\documentclass{mem}
\usepackage{natbib}\usepackage{txfonts}\usepackage{balance}
\usepackage{graphicx}
\usepackage[a4paper]{hyperref}
\idline{75}{282}

\begin{document}

\def\teff{$T\rm_{eff }$}
\def\kms{$\mathrm {km s}^{-1}$}
\def\beq{\begin{equation}}
\def\eeq{\end{equation}}


\title{Self-Regulation of Solar Coronal Heating
via the Collisionless Reconnection Condition}

\author{D. \,Uzdensky\inst{1}}

\offprints{D. \,Uzdensky}

\institute{Dept. of Astrophysical Sciences, Princeton University,
Princeton, NJ 08544, USA, \\ and Center for Magnetic Self-Organization;
\ \email{uzdensky@astro.princeton.edu}
} 

\authorrunning{Uzdensky}

\titlerunning{Solar Coronal Heating}

\date{Feb. 9, 2007}

\abstract{
I present a novel view on the problem of solar coronal heating.
In my picture, coronal heating should be viewed as a self-regulating 
process that works to keep the coronal plasma marginally collisionless.
The self-regulating mechanism is based on the interplay between two effects:
(1) Plasma density controls coronal energy release via the transition 
between the slow collisional Sweet-Parker regime and the fast collisionless 
reconnection regime;
(2) In turn, coronal energy release through reconnection leads 
to an increase in the ambient plasma density via chromospheric
evaporation, which temporarily shuts off any subsequent reconnection 
involving the newly-reconnected loops.
}

\maketitle{}


I here discuss certain aspects of solar coronal heating 
[see \cite{klimchuk2006} for a recent review] in the context of 
the \cite{parker1988} nano-flare model. Since the main heating 
process in that model is magnetic reconnection, I will first 
discuss what we have learned about reconnection in the past 20 years. 
Even though we still don't have a complete picture of reconnection, 
there is now consensus about some of its fundamental aspects. 
My main goal is to use this emerging knowledge to shed some new light 
on the old coronal heating problem.

First, I would like to emphasize the importance of a realization by 
\cite{petschek1964} that the main bottleneck in the classical Sweet--Parker
\citep{sweet1958,parker1957} reconnection model is the need to have 
a reconnection layer that is both thin enough for the resistivity 
to be important and thick enough for the plasma to be able to flow out. 
Furthermore, \cite{petschek1964} proposed that this can be resolved if 
the reconnection region has a certain special structure: the Petschek 
configuration, with four shocks attached to a central diffusion region. 
Then, there is an additional geometric factor that leads to faster 
reconnection. 
This idea is especially important in astrophysical systems, including 
the solar corona, irrespective of the actual microphysics inside the layer. 
This is  because the system size~$L$ is much larger than any microscopic 
physical scale~$\delta$, e.g., the ion gyro-radius~$\rho_i$, the ion 
collisionless skin-depth~$d_i \equiv c/\omega_{pi}$, or the Sweet--Parker 
layer thickness~$\delta_{\rm SP}=\sqrt{L\eta/V_A}$. Then, a simple 
Sweet--Parker-like analysis would give a reconnection rate $v_{\rm rec}/V_A$ 
scaling as~$\delta/L \ll 1$, and hence would not be rapid enough 
to be of any practical interest. 
Thus, we come to {\bf Conclusion~I}: {\it Petschek's mechanism (or its 
variation) is necessary for sufficiently fast large-scale reconnection.}

However, several numerical and analytical studies 
[e.g., \cite{biskamp1986,scholer1989,uk2000,kulsrud2001,mlk2005}] 
and even laboratory experiments \citep{ji1998}
have shown that in resistive MHD with a uniform (and, 
by implication, Spitzer) resistivity Petschek's mechanism does not work 
and the slow Sweet--Parker scaling applies instead. 
Thus, we come to {\bf Conclusion~II}: 
{\it In the collisional regime, when classical resistive MHD applies, 
one does not get Petschek reconnection.}

The natural question to ask now is whether fast reconnection possible 
in a collisionless plasma where resistive MHD doesn't apply.
There is a growing consensus that the answer is YES. 
First, in space and solar physics there has long been a serious evidence 
for fast collisionless reconnection; recently it has also been confirmed 
in laboratory studies \citep{ji1998,yamada2006}. 
At the same time, several theoretical and numerical studies have 
recently indicated that fast reconnection, enhanced by the Petschek 
mechanism, does indeed take place in the collisionless regime. 
Moreover, it appears that there are even two physically-distinct 
mechanisms for fast collisionless reconnection:
{\bf (i)} {\it Hall effect} [e.g., \cite{shay1998,gem2001,bmw2001,cassak2005}];
and 
{\bf (ii)} spatially-localized {\it anomalous resistivity}
[e.g., \cite{ut1977,sh1979,scholer1989,kulsrud2001,bs2001,mlk2005}].
At present, we still don't know which of these two mechanisms operates 
under which circumstances and how they interact with each other. However, 
they both seem to work and both seem to involve an enhancement due to a 
Petschek-like configuration. Thus, we can draw {\bf Conclusion~III}:
{\it a Petschek-enhanced fast reconnection does happen in the collisionless
regime.}

To sum up, there are two regimes of magnetic reconnection:
slow Sweet--Parker reconnection in resistive-MHD with classical 
collisional resistivity, and fast Petschek-like reconnection in 
collisionless plasmas.

How can one quantify the transition between the two regimes?
Consider for simplicity the case with no guide field $B_{\rm guide}=0$, 
(If $B_{\rm guide}\neq 0$, some of the arguments and results presented
below may be modified, but they will remain conceptually similar). 
Then, the condition for fast collisionless reconnection can be formulated 
[e.g., \cite{kulsrud2001,uzd2003,cassak2005,yamada2006}] roughly as 
\beq
\delta_{\rm SP} < d_i , .
\label{eq-1}
\eeq
Expressing resistivity in terms of the Coulomb-collision electron mean-free 
path~$\lambda_{e,\rm mfp}$, one can write \citep{yamada2006}:
\beq
{{\delta_{\rm SP}}\over{d_i}} \sim 
\biggl({L\over{\lambda_{e,\rm mfp}}}\biggr)^{1/2}\ 
\biggl(\,{m_e\over{m_i}}\biggr)^{1/4} \, .
\label{eq-2}
\eeq
where I have neglected numerical factors of order~1 and used
the condition of force balance between the plasma pressure ($2 n_e T_e$)
inside and the reconnecting field pressure ($B_0^2/8\pi$) 
outside the layer. 
Then, the above fast reconnection condition becomes
\beq
L < L_c \equiv \sqrt{m_i/m_e}\, \lambda_{e,\rm mfp} 
\simeq 40 \,  \lambda_{e,\rm mfp} 
\label{eq-3}
\eeq

The mean-free path is given by $\lambda_{e,\rm mfp} \simeq 
7\cdot 10^{7}{\rm cm}\, n_{10}^{-1}\, T_7^2 $, 
where we set $\log\Lambda \simeq 20$ and where~$n_{10}$ and~$T_7$ 
are the central layer density~$n_e$ and temperature~$T_e$ in units 
of $10^{10}\, {\rm cm}^{-3}$ and~$10^7$~K, respectively. 
Then equation (..)  becomes:
$L < L_c(n,T) \simeq 3\cdot 10^{9}{\rm cm}\, n_{10}^{-1}\, T_7^2 $. 
The strong temperature dependence indicates that knowing~$T_e$ 
at the center of a Sweet--Parker layer is crucial. If there is 
no guide field,~$T_e$ follows readily from the cross-layer pressure 
balance: $T_e \simeq 1.4 \cdot 10^7\, {\rm K}\ B_{1.5}^2\, n_{10}^{-1} $ 
[here $B_{1.5}\equiv B_0/(30\,{\rm G})$]. Moreover, even if $B_{\rm guide}
\neq 0$, one can show that this estimate still approximately holds. 
As a result, the collisionless reconnection condition becomes \citep{uzd2006} 
\beq
L < L_c(n,B_0) \simeq 6\cdot 10^{9}\,{\rm cm}\ n_{10}^{-3}\, B_{1.5}^4 
\label{eq-4}
\eeq


Let us now discuss the implications of these results for the solar corona. 
I propose that {\it coronal heating is a self-regulating process 
keeping the corona marginally collisionless} in the sense of 
equations~(\ref{eq-3})-(\ref{eq-4}) [see \cite{uzd2006}].

As long as flux emergence and the braiding of coronal loops by photospheric 
footpoint motions keep producing current sheets in the corona, magnetic 
dissipation in these current sheets results in intermittent coronal heating 
\citep{rtv1978,parker1988}. 
Typical values of~$L$ and~$B_0$ of these current sheets are determined by 
the emerging magnetic structures and by the footpoint motions. Therefore, 
here I will regard~$L$ and~$B_0$ as known and ask what determines the 
coronal density and temperature.

Resolving~(\ref{eq-4}) with respect to~$n_e$, we get a critical density, 
$n_c$, below which reconnection switches from the slow collisional regime to 
the fast collisionless regime:
\beq
n_c \sim 2\cdot 10^{10}\, {\rm cm}^{-3}\, B_{1.5}^{4/3}\, L_9^{-1/3} \, .
\label{eq-5}
\eeq
(Here $L_9=L/10^9\, {\rm cm}$.) This value is close to that observed 
in active solar corona. I suggest that this is not a coincidence.

As an example, consider a coronal current sheet with some~$L$ and~$B_0$. 
If initially the ambient density~$n_e$ is higher than~$n_c(L,B_0)$, the 
current layer is collisional and reconnection is in the slow mode. 
Energy dissipation is weak; the plasma gradually cools radiatively 
and precipitates to the surface. The density drops and at some point 
becomes lower than~$n_c$. Then, the system switches to the fast 
collisionless regime and the rate of magnetic dissipation jumps. 
Next, there is an important positive feedback between coronal energy 
release and the density. A part of the released energy is conducted
to the surface and deposited in a dense photospheric plasma. This leads 
to chromospheric evaporation along the post-reconnected magnetic loops. 
As a result, these loops become filled with a dense and hot plasma. 
The density rises and may now exceed~$n_c$. This will shut off any
further reconnection (and hence heating) involving these loops until
they again cool down, which occurs on a longer, radiative timescale.

Thus we see that, although highly intermittent and inhomogeneous, 
the corona is working to keep itself roughly at the height-dependent 
critical density given by equation~(\ref{eq-5}). In this sense, coronal 
heating is a self-regulating process \citep{uzd2006}.

\begin{acknowledgements}

This work is supported by National Science Foundation Grant 
No.\, PHY-0215581 (PFC: Center for Magnetic Self-Organization 
in Laboratory and Astrophysical Plasmas).

\end{acknowledgements}

\bibliographystyle{aa}

\begin{thebibliography}{}

\bibitem[{Bhattacharjee et al. (2001)}]{bmw2001}
Bhattacharjee, A., Ma, Z.\ W., \& Wang, X.\ 2001, Phys.\ Plasmas, 8, 1829

\bibitem[{Birn et al. (2001)}]{gem2001}
Birn, J.\ et al.\ 2001, \jgr, 106, 3715

\bibitem[{Biskamp (1986)}]{biskamp1986}
Biskamp, D.\ 1986, Phys. Fluids, 29, 1520.

\bibitem[{Biskamp \& Schwarz (2001)}]{bs2001}
Biskamp, D., \& Schwarz, E.\ 2001, Phys.\ Plasmas, 8, 4729

\bibitem[{Cassak et al. (2005)}]{cassak2005}
Cassak, P., Shay, M., \& Drake, J.\ 2005, \prl, 95, 235002

\bibitem[{Ji et al. (1998)}]{ji1998}
Ji, H., et al.\ 1998, \prl, 80, 3256

\bibitem[{Klimchuk (2006)}]{klimchuk2006}
Klimchuk, J.~A.\ 2006, Solar Phys., 234, 41

\bibitem[{Kulsrud (2001)}]{kulsrud2001}
Kulsrud, R.~M.\ 2001, Earth, Planets and Space, 53, 417

\bibitem[{Malyshkin et al.\ (2005)}]{mlk2005}
Malyshkin, L.~M., Linde, T., \& Kulsrud, R.~M.\ 2005, 
Phys. Plasmas, 12, 102902

\bibitem[{Parker (1957)}]{parker1957}
Parker, E.~N.\ 1957, \jgr, 62, 509

\bibitem[{Parker (1988)}]{parker1988}
Parker, E.~N.\ 1988, \apj, 330, 474

\bibitem[{Petschek (1964)}]{petschek1964}
Petschek, H.~E.\ 1964, AAS-NASA Symposium on Solar Flares, 
(National Aeronautics and Space Administration, Washington, 
DC, 1964), NASA SP50, 425.

\bibitem[{Rosner et al. (1978)}]{rtv1978}
Rosner, R., Tucker, W.~H., \& Vaiana, G.~S.\ 1978, \apj, 220, 643

\bibitem[{Sato \& Hayashi (1979)}]{sh1979}	
Sato, T.\ \& Hayashi, T.\ 1979, Phys. Fluids, 22, 1189

\bibitem[{Scholer (1989)}]{scholer1989}
Scholer, M.\ 1989, \jgr, 94, 8805.

\bibitem[{Shay et al. (1998)}]{shay1998}
Shay, M.~A., et al.  1998, \jgr, 103, 9165

\bibitem[{Sweet (1958)}]{sweet1958}
Sweet, P.~A.\ 1958, in IAU Symp.~6, Electromagnetic Phenomena in Cosmical 
Physics, ed.\ B.~Lehnert, (Cambridge: Cambridge Univ. Press), 123.

\bibitem[{Ugai \& Tsuda (1977)}]{ut1977}
Ugai, M., \& Tsuda, T.\ 1977, J. Plasma Phys., 17, 337.

\bibitem[{Uzdensky \& Kulsrud (2000)}]{uk2000}
Uzdensky, D.~A., \& Kulsrud, R.~M.\ 2000, Phys. Plasmas, 7, 4018.

\bibitem[{Uzdensky (2003)}]{uzd2003}
Uzdensky, D.~A.\ 2003,\apj, 587, 450

\bibitem[{Uzdensky (2006)}]{uzd2006}
Uzdensky, D.~A.\ 2006, ArXiv Astrophysics e-prints, astro-ph/0607656

\bibitem[{Yamada et al. (2006)}]{yamada2006}
Yamada, M., et al.\  2006, Phys. Plasmas, 13, 052119




\end{thebibliography}
{}

\end{document}